# The Wall and The Ball:
# A Study of Domain Referent Spreadsheet Errors

Richard J. Irons
Faculty of Business and Law
Central Queensland University
Rockhampton Qld. Australia.

**ABSTRACT**

*The Cell Error Rate in simple spreadsheets averages about 2% to 5%. This CER has been measured in domain free environments. This paper compares the CERs occurring in domain free and applied domain tasks. The applied domain task requires the application of simple linear algebra to a costing problem. The results show that domain referent knowledge influences participants' approaches to spreadsheet creation and spreadsheet usage. The conclusion is that spreadsheet error making is influenced by domain knowledge and domain perception. Qualitative findings also suggest that spreadsheet error making is a part of overall human behaviour, and ought to be analyzed against this wider canvas.*

## 1 INTRODUCTION

Spreadsheet errors are pervasive, stubborn, ubiquitous and complex. Research over the last twenty years has shown them to be present in nearly all spreadsheets, with an incidence rate of about 2 to 5 % of all non-text cells. This research has proceeded on several fronts. One approach has been the identifying, classifying, and measuring of errors. Another has been the establishment of educational regimes which might reduce error making, whilst another has aimed to develop auditing tools which will help uncover errors. The overall aim has been the eradication of errors. Eradication is essential if businesses are to make reliable decisions based on spreadsheet outcomes.

Despite the most strenuous corrective exhortations from writers, spreadsheet errors still remain. For example, Schlosser in an excellent 1989 article gives checklists of possible errors, including a pervasive classic:

"a variable which is not defined as an 'if statement when it should be (this is an surprisingly frequent mistake with taxes)"

The pervasiveness of this error is demonstrated in a reputable text published nine years later:

| Year | 0 | 1 | 2 |
|---|---|---|---|
| ***Income statement*** | | | |
| Sales | $1,000 | $1,070 | $1,145 |
| Costs of goods sold | | (910) | (973) |
| Interest payments | | (33) | (34) |
| Depreciation | | (123) | (142) |
| Profit before tax | | 5 | (5) |
| Taxes | | (2) | 2 |
| Profit after tax | | 3 | (3) |
| Dividends | | (2) | 2 |
| Retained earnings | | 1 | (1) |

Table 1: Part of Spreadsheet table 'PRO FORMA 4',

Benninga, S.,(1997), p 13.

This spreadsheet demonstrates the calculation of a firm's target debt/equity ratio. It is the same spreadsheet introduced at the start of the chapter which is used to demonstrate financial statement forecasting. There is no 'if statement to check whether tax is being paid on negative accounting income. Hence, in year two, positive tax is paid on negative income. Similarly, positive dividends are being paid on the same negative income. A 'simple' logic trap would have avoided these errors. This error remains even in the second edition of the text.

As this error is 'obvious' there is a suspicion that the code writer has not audited the model. This is not only the writer's problem, but it is also a problem for those contributors who have advised corrections for the second edition of the text to the website at http://finnace.wharton.upenn.edu/-benninga/ This website invites correction to the text from readers, and lists discovered errata. It is a good idea, and uses the power of worldwide communication to enhance a single piece of work. The contributions to the website cover both book text errors, and Excel errors. The Excel errors noted cover the 'usual' range of spreadsheet errors; mis-keying, wrong cell references, and errors in formulae. None of the errata messages mention the omission of the noted 'if statement. These exchanges beg the question: are these errors sighted because finance referent knowledge is different from accounting referent knowledge, or has the error just been 'overlooked'?

Thus the question of domain knowledge arises. One would expect that an author who sets out to write a text on financial modeling, is expert in accounting, economics, finance and the requisite tools of algebra and statistics. Hence, there is a presumed suite of domain knowledges. However, this presumption may be unfounded. Accounting is not the same as finance; finance is not the same as economics. More importantly, code writing is an art in itself, and it must be combined with domain knowledge to be useful in a computer modeling setting. The underlying formulae in this spreadsheet demonstrates this idea:

| A | B | C | D | E |
|---|---|---|---|---|
| | **Year** | **0** | **1** | **2** |
| **16** | **Income statement** | | | |
| **17** | Sales | =B3 | =B17*(1+$B$4) | =C17*(1+$B$4) |
| **18** | Costs of goods sold | | =-C17*$B$8 | =-D17*$B$8 |
| **19** | Interest payments | | =-$B$10*C37 | =-$B$10*D37 |
| **20** | Depreciation | | =(C32-B32) | =(D32-C32) |
| **21** | Profit before tax | | =SUM(C17:C20) | =SUM(D17:D20) |
| **22** | Taxes | | =-C21*$B$11 | =-D21*$B$11 |
| **23** | Profit after tax | | =C22+C21 | =D22+D21 |
| **24** | Dividends | | =-C23*$B$13 | =-D23*$B$13 |
| **25** | Retained earnings | | =C24+C23 | =D24+D23 |

Table 2: Formula view of spreadsheet shown in Table 1.

Taxes are computed in row 22 as: -C21 *$B$1 1 etc, where B 11 holds the tax rate. To compute taxes as an expected negative, the value of 'profit before tax' from row 21 is artificially made negative. However, if profit before tax in row 21 is already negative, then taxes will be calculated as a positive amount, and will be paid on negative income. The fail safe method would have been to create an 'if, then' logic trap in row 22 to prevent the payment of taxes on negative income. (If carry forward of tax losses is permitted, then more extensive logic is required). A similar trap could have been employed in row 24 to prevent the payment of dividends from negative income.

Why have artificial negatives been created in rows 22 and 24? Without second guessing the code writer, it might be argued that this method is simple, 'quick and dirty' and is a simplistically sufficient for the demonstration at hand. That is, taxes and dividends will always be negative outflows, so the code is set up to generate those flows without extensive formulae. On the other hand, the code writer may have not had sufficient domain knowledge to realize that taxes and dividends are not paid on negative income.

Calculation of taxes and dividends relies on accounting domain knowledge. The notion of cash flow within a computerized budget comes from finance. Given reasonable world- wide curricula, one would expect that accounting and finance knowledge would be in the same domain, but this idea cannot be taken for granted. Additionally, the code writer may have simply ignored the fact that income can be negative at times, or alternatively, lacking domain knowledge, might have assumed that 'income' can never be negative.

Whatever the case, poor code writing has led to an obvious error. The point is that there may be gap between the domain knowledge actually held by a code writer, and the level of domain knowledge that the model requester expects the code writer to have. This gap may be a factor in spreadsheet error making.

## 2 THE RESEARCH VEHICLE

One task set up to investigate error making in a domain free environment is Panko and Sprauge's (1998) (PS) 'Wall' task. The only domain knowledges ostensibly required in the task are simple arithmetic and some concept of costs and profits. It could be expect that these are widely held life knowledge domains coming from a high school education. In fact the authors reported that only 5% of participants in their experiment regarded the task as 'difficult'.

The current experiment set out to investigate the extent of the gap between actual and expected domain knowledges.

## 3 THE RESEARCH METHOD

The research set out to investigate three main themes:

1. The frequency of errors in end user created spreadsheets
2. Whether the difficulty of the spread sheeting task has any influence on the frequency of errors
3. The attitudes participants held toward spreadsheet errors.

The research was carried out by asking volunteers to prepared two separate spreadsheets to solve two computational tasks. These were the PS Wall task, and the Ball task. (The Ball task is similar in style to the Wall task, but with an increased algebraic requirement). Participants were asked to complete each of these tasks in their own time at their own pace, and were asked not to seek outside help. The completed spreadsheets were reviewed by the researcher and graded for errors. Following this grading, the participants were invited to a one on one interview which took about 10 minutes. This interview used a structured questionnaire, which was partially self administering, and partially interactive. The task completion-interview approach combined the methodologies of Nardi and Miller (1991) and Hendry and Green (1994).

The questions were framed to discover the experience and skill backgrounds of the participant, and then to seek the participant's attitude to the discovered errors. Attitude could be an important variable in error research, as Brown and Gould (1987) found that subjects' confidence in completed spreadsheets was often unfounded.

The questionnaire was designed so that subjects were moved progressively from the general notion of spread sheeting through to the particular investigation of actual self made errors. The questionnaire is attached as appendix A.

When the "Cell Error Rate" is used as the error frequency measure, the average reported CER is about 5%. Panko and Halverson (1996) reported a range of 1.7% to 9.3% in a survey of studies on Cell Error Rates. In most of these studies, the CER was measured over non-text cells.

However, in both the current Wall and The Ball tasks, cells containing text, such as title cells and heading cells were included in the error count analysis. Text cells were included because an error in these cells could have a bearing on any final decision made using the spreadsheet output. In the particular case here, the author inadvertently reversed the column names on the solution for the Ball task. Had this error not been identified, a decision maker could have been left with the wrong answer for the Hydrogen and Helium alternatives. This is not an error of computation, but would have become an error of decision making.

Therefore, it may be necessary to review the standard approach in the literature which generally excludes text information from CER counts. This paper introduces the term OCER to measure the Overall Cell Error Rate. The OCER uses **all** cells in a spreadsheet.

## 4 THE TASKS

The Wall task is a direct copy of the task used by Panko and Sprauge (1998). The Wall task asks participants to calculate the cost of combining labour and material costs in the construction of a garden wall. There are two materials, lava rock or brick, which allow some testing of different costing calculations. The only change made to this task for the current study was the change of measurement from feet to meters for a local audience. Panko and Sprauge (1998) argue that the Wall task focuses more directly on the process of spread sheeting as the only domain knowledge required is the volumetric calculation for the rectangular solid.

The Ball task was designed to be computationally more difficult, as it required application of algebra to manipulate spherical volume and area, and to manipulate gas volume and pressure relationships under Boyle's Law. Not only did the problem visually appear more difficult it required careful application of domain knowledge. Some guidance as to how to apply the domain knowledge of Boyle's Law was given in the task description.

The full texts of the Wall and The Ball tasks appear as appendices B and C respectively.

## 5 THE PARTICEPANTS

Participant volunteers were sought from amongst undergraduate full time students and from academic staff colleagues. Unfortunately, the uptake rate from both groups was quite low. Only one student from a class of about 60 followed the experiment through. In a second class of about 10 students, only 3 students volunteered for the experiment, but only one of these completed all aspects in full. Six academic colleagues completed the experiment out of a potential group of about 20. One professional colleague completed all tasks. This gave eleven participants in all. Some of the participants had to be vigorously pursued to complete the tasks.

The low participation rates, and poor adherence to timelines, suggests that volunteering is not popular, especially when there is no incentive reward. Advice from colleagues differed on this issue. Some argued that a small incentive, such a book voucher, would be useful, whilst others argued that this scheme would bias the responses. Panko and Sprauge (1998) offered course credit for participation, but that option was not open in the current circumstance. It would appear that some form of reward is necessary to secure adequate participation rates.

## 6 THE PROCESS

Volunteer subjects were given the tasks, and were asked to complete them in their own time, to a standard which they themselves considered "reasonable". They were not to seek outside help, and were advised to spend say not more than 20 minutes on each task. There were no definite time lines set for the end of the experiment but the expectation was that the tasks would be completed, and the subsequent interviews conducted within two weeks of subjects receiving the material. As it turned out, there was a large variation in completion time of each individual participant. The shortest completion time was about one week, whilst the longest was about 20 weeks. It was difficult to draw any conclusions from these varying time lengths, but it would appear natural that a long time between task completion and questionnaire completion would be best avoided.

## 7 ERROR MEASUREMENT PROTOCOLS

Panko and Sprauge (1998) highlight the difficulty of arriving at a particular definition of the CEP, and on the added difficulties in classifying the errors. In their paper they chose a three part classification scheme: mechanical errors, logic errors and omission errors, following Allwood (1984).

Mechanical errors include mis-keying, pointing to the wrong cell, reading in a wrong number and selecting the wrong range.

Logic errors include faulty reasoning, using the wrong algorithm or implementing the algorithm with the wrong logic.

Omission errors simply mean the leaving out of something from a model.

Other writers have further subdivided or amalgamated these schemes. For example, Panko and Halverson (1997), (PH), subdivided logic errors in to Eureka errors; those which are easy to prove to be incorrect, and Cassandra errors; those which are difficult to prove to be incorrect.

A more comprehensive taxonomy has been proposed by Rajalingham, Chadwick and Knight (2001), (RCK). This taxonomy is developed by progressive binary division from an initial binary classification. The initial classification is "system generated errors" and "user generated errors". The final branches of the binary tree contain numerous classifications.

The detail in the P H and RCK schemes will be useful for long term research into research prevention. For the present study they are too detailed. The present study focused on replicating the original PH Wall study, and in extending this into an applied domain knowledge application.

In this present study, only the coarsest classification of errors was used. Only three error types were identified. These were: readability, logic algebra.

Readability is a subjective error identifier. Whether a spreadsheet is high or low in readability depends upon the expectations of the reader. If the reader wants only a bottom line total, then readability is of little concern. However if the reader wants more detailed information, such as subtotals for material costs, labour costs, and profit markup in the wall project, then the omission of such items creates readability errors. Secondly, the layout of the results may make it difficult for a reader to discern the required information. Thirdly, readability enhancements such as the use of fonts, highlighting and colour may be missing, inappropriately applied, or applied to the wrong sections of the results.

The question of readability is also domain dependent. Within a business managerial environment there may be some implicit or tacit understanding that managers require a printed report showing computational detail, progressive calculations, or segmented outputs. These attributes may enhance confidence in decision-making, add to a manger's grasp of the problem, or allow decisions on sub-parts of problems. Gross bottom line figures may be correct in certain instances, but they may not rest easy with managers. Hence, readability is a key issue within a decision environment. Additionally, the function of spreadsheet is that output is presented as a report, not as a summary calculation. If only a summary calculation is required, then a hand held calculator will often provide it.

Readability ought to be an issue in spreadsheet error research, regardless of its subjective nature. Its subjectivity is vitally important though in establishing an error rate. Overall "readability" cannot be attributed to a single cell, or single cells. So as to include readability into a CER measure, some *ad hoc* counting system must be used. In the present case, a notional number of cell errors were allocated, depending upon the implied information that management may require. Whilst this is an unsatisfactory measure, it allows for a quantifiable result.

Logic error in the present study was a catchall term for errors in reasoning, errors in data relationships, and for all other keying, pointing and relationship procedures. This is a coarse measure, but as the study here investigated overall error counts then this catchall classification was satisfactory. The PS study divided this overall classification into omission, logic and mechanical errors.

A separate category of errors was established for the present study. This classification was called "algebra". The hypothesis leading to the creation of this class was that in moving from the Wall to the Ball tasks, participants were likely to make more errors in manipulating the algebraic formulae for the spherical measures and the gas measures. The 'algebra' category of errors was designed to catch these errors. In the PS study, this special class would have been included in the overall "logic" class.

## 8 ERROR MEASUREMIENT PARAMETERS

The current study uses two definitions of the CER. One is a wide definition which includes **all** cells in each spreadsheet. This global inclusion allows for measures of readability, by assuming that all cells will have some impact upon a decision maker. Therefore, cells containing headings, titles, text and other information outside of the narrow "computational" definition have been included in this measure. The global measure to capture errors in **all** cells is termed the Overall Cell Error Rate, OCER. For direct comparison with the PS results, the standard CER is also measured. This is constructed by adding the errors in the "logic' and "algebra" classifications.

## 9 THE RESULTS

Summary statistics for the experiment are given below.

**Spreadsheet Error Analysis: Summary Statistics**

| | The WALL | The BALL |
|---|---|---|
| Total Participants | 11 | 11 |
| Total Submitted S'Sheets | 11 | 7 |
| S'Sheets With Correct Bottom Line | 9 | 2 |
| S'Sheets With Wrong Bottom Line | 2 | 5 |
| % of S'Sheets with Errors | 18.18% | 71.42% |
| % Useful for Decision Making | 72.72% | 28.57% |
| Average CER | 1.67% | 11.86% |
| Average OCER | 6.9% | 12.86% |

Table 3: Summary statistics for the Wall and the Ball experimental tasks.

Results from these measures are:

The Wall CER rate of 1.67 % is **not** significantly different from zero at the 5% level. The value is at the lowest level of the Panko and Sprauge range. Nine spreadsheets out of 11 showed no errors. The two spreadsheets with errors had multiple errors. These two results were actually extreme outliers in an otherwise perfect completion rate.

The Ball CER rate of 11. 86 % is significantly different from zero at the 5% level.

The Wall CER at 1.67% and the Ball CER at 11.86% are **not** significantly different from each other at the 5% level.

Both the Wall OCER at 6.9% and the Ball OCER at 12.86% **are** significantly different from zero at the 5% level.

The Wall OCER at 6.9% and the Ball OCER at 12.86% are not significantly different from each other at the 5% level.

The percent of spreadsheets with errors is similar to the Panko and Sprauge findings.

## 10 COMMENTS ON RESULTS.

The Wall task was set out first to give participants some practice in the type of calculation required. The Ball was set out second, as it required arguably more difficult calculations. Participants were not required to do the tasks in that order, but the interviews revealed that no one did the tasks in the reverse order.

The difference in the number of submitted Wall and Ball spreadsheets indicates that several participants found the Ball task simply 'too hard'.

Where both spreadsheets were submitted there was a difference in the number of errors between both tasks. However, this difference is not statistically significant. What is interesting, but not statistically tested, is the finding that the percentage of useful Ball spreadsheets is much less than the percentage of useful Wall spreadsheets.

Interestingly, some participants got the Ball correct, but made simple errors on the Wall. In other cases, some participants had trouble following the written instructions especially with respect to the final balloon volume and the related measures of waste. Some participants attempted to, but could not manipulate Boyle's law. Some participants simply made algebraic errors in manipulating the area and volume formulae.

## 11 COMMENTS ON DOMAIN KNOWLEDGE

The results show that participants had trouble in both making an attempt at, and in completing, the Ball task. Those who did not attempt it were unanimous is saying that it was 'too hard'. Most of those who attempted it, it the confident understanding that they could solve it, made errors in both logic and algebra. Errors in logic were in interpreting the initial data, structuring the calculation, and in working to a result. Errors in algebra were caused by poor manipulation of the spherical measures, and in the gas relationships.

Domain referent knowledge is more apparent if we examine the correlation between the two domain areas. The correlation is based on bottom line results.

**Domain Knowledge Matrix**

| Wall / Ball | Participants |
|---|---|
| Right / Right | 1 |
| Right / Wrong | 8 |
| Wrong / Right | 1 |
| Wrong / Wrong | 1 |

Table 4. Number of participants who achieved a particular mix of solutions.

Includes those participants who did not finish tasks.

Interestingly, only one participant completed both tasks correctly, whilst 8 participants got the Wall correct and the Ball wrong. An anomalous result is that one participant got the 'easy' Wall task wrong, and the 'hard' Ball task correct. This particular participant treated the two tasks as interesting intellectual challenges, and hurried through the simple Wall task in order to tackle the more interesting Ball task.

The eight 'right/wrong' mix results highlight the implication that domain knowledge is an important variable in error investigation.

There are several other conclusions to draw from both the actual spreadsheet results, and from the interview responses:

- Those who attempted the Ball thought that its required level of domain knowledge was a reasonable expectation : ' ... it required only high school algebra.'

- Others attempted it, but could not solve key parts, the calculation of the radius for example. One participant manipulated the algebra fully, leaving r^3 as unsolved, as he could not calculate '... the cubic square root.'; ( he understood the general notion of a cube root, but was not sure how to solve it or how to use an Excel function to solve it.)

- Other participants thought that it was an interesting intellectual challenge, and rushed to get a solution, only to be disappointed when their result was wrong.

## 12 COMMENTS ON SPREADSHEET APPLICATIONS GENERALLY

Research into spreadsheet error is motivated by the desire to prevent dangerous managerial decision making. Spreadsheets are a powerful computational and communication tool, and there is some inherent danger in that a poorly prepared or audited spreadsheet could be naively accepted by decision makers as providing reliable information. To test whether the attractiveness and apparent simplicity of spread sheeting could lead to such problems, participants were asked;

'Do you think spreadsheet errors might arise because spread sheeting encourages inexperienced users?'

Some respondents agreed with this statement, but most respondents stated that errors were bound to occur, but that the attractiveness of spread sheeting as not necessarily the cause. Whilst the ease and apparent facility of spread sheeting was an issue, the bigger issues identified were lack of training, lack of error detection concepts, and lack of checking. On the other hand, those respondents who lacked confidence to attempt the Ball, stated that they would seek help with such difficult tasks, rather than embarking upon them in ignorance.

Overall, the outcomes of the interviews suggest that under-confidence is a natural preventative in tackling difficult tasks, and that overconfidence is a serious precursor to spreadsheet error making.

Other issues which arose and shed some light on how errors are made were;

- Only one respondent submitted a pencil and paper attempt, even though this was requested in the instructions

- Even though the Ball and Wall tasks lent themselves to visual representation, no respondents sketched the relative pictures to aid in envisioning the sizes and volumes of the structures to help in making reasonable calculations.

Hence, despite the clear and early exhortations of Schlosser (1998), participants are still rushing to the computer to get an early solution, without firstly making a pencil and paper estimate.

## 13 CONCLUSION

Domain knowledge is an important variable in spreadsheet error making. There is a gap between the domain knowledge expected of a modeler, and that actually held by a modeler.

Errors also arise because of ordinary human traits, such as: procrastination, lack of interest, lack of care or lack of attention to the task. Additionally, overconfidence and under-confidence can lead to errors or task incompletion. Apparently, spreadsheet errors need to be portrayed against the wider canvas of basic human behavioural norms.

It is going to be difficult to continue to address error making as only an impersonal, technocratic process. Managing the human condition seems to be the long term solution to spreadsheet error making.

Potential questions which remain for further research are:

- Can error making be properly tested in voluntary laboratory settings, where the outcomes may be more a function of time and task commitment than simply spread sheeting skill?

- Is the Ball task a true test of spreadsheet error making, or would the same errors have been made when participants used pencil and paper. In other words, is spread sheeting or algebra being tested here?

- Given that some participants were indifferent to the discovered errors, or were tardy in completing tasks, would these attitudes be carried over to real world work places to such a degree that mechanical error checking processes would be ignored?

An interesting research future remains.

# Appendix A.
# Errors in Spreadsheets Questionnaire:
# 'The Ball' and, 'The Wall' Tasks

**Responses are to be entered by the interviewer. More than one response is applicable in some cases Please place a tick in the box where appropriate.**

**1 How would you rate your level of experience with spreadsheets?**

High Moderate Low None

**2 How long have you been involved in the use of spreadsheets?**

0-1 years 2-5 years 6-10 years more than 10 years

(Where possible, specify a number of years …………………………………………………..)

**3 Describe your use of spreadsheets. (Tick those which are applicable.)**

Creation

Creation and Use

Auditing Spreadsheets of Others

Data entry to spreadsheets

Using spreadsheet output

**4 What would be the average size, in number of cells, of spreadsheets you are most familiar with? (The number of cells is the product of the number of rows and columns, used for computation.)**

100 400 2,500 more than 10,000

**5 When dealing with the size of spreadsheet you are familiar with, how confident are you of The accuracy of the spreadsheet results?**

Absolutely definite Reasonably confident Uncertain Very unsure

**6 What activity do you think would cause the greatest number of errors within any spreadsheet? (Select one response only.)**

Mistyping of values Incorrect cell referencing Incorrect formulae

Incorrect problem definition Ignorance of spreadsheet operation

**7 How confident are you in the results given by the spreadsheets you have created for the Wall and the Ball tasks?**

Not at all confident Mildy unsure Reasonably confident Absolutely certain

**8 How many computational errors do you think you have created in the Wall and the Ball spreadsheets. (Ignore headings and text errors.)**

**The Wall Task**

0 5 10 more than 10

**The Ball Task**

0 5 10 more than 10

**9 The number of errors actually discovered in each of your spreadsheets is:**

**The Wall** ……………….

**The Ball** ……………….

**10 For each of the following errors, can you suggest why the error was made.**

| Error Type | Possible Reason (Select From the lists below.) |
|---|---|
| | |
| | |
| | |
| | |
| | |
| | |
| | |
| | |

- **Software Usage errors:-** Keying; layout; data omission; understanding of task; logic structure; use of algebra; formulae structure; spreadsheet function; spreadsheet knowledge.

- **Behavioural errors:-** Fatigue; time pressure; task importance; interest; consequences; task size; task knowledge; background influences; confidence.

**11 Suggest ways in which you might overcome these errors in future.**

| Error Type | Possible Solutions (Select from the lists below.) |
|---|---|
| | |
| | |
| | |
| | |
| | |
| | |
| | |
| | |
| | |

- **Software solutions:-** More time; more commitment; more spreadsheet training more familiarity with problem; discussion with colleagues; working in groups; applying error check routines; second person audit; worked template; known solution.

- **Behavioural solutions:-** Ensuring mental freshness; more confidence in task ability; more empathic support; more experience; smaller task; more training in using spreadsheets.

**12 "Errors can be made in spreadsheets". How might this statement affect your future use of spreadsheets?**

- **Possible effects:** Cause more time to be spent in checking spreadsheets; spend more time in training users and creators; introduce more errors finding software techniques; duplicate spreadsheet creating; expensively audit all spreadsheets; use spreadsheets as a guide only - not a definitive solutions; avoid spreadsheets altogether - use some other form of computation.

**13 Give your views on the following: "Do you think spreadsheet errors might arise because spread sheeting encourages inexperienced users?"**

..................................................................................................................................................................

..................................................................................................................................................................

..................................................................................................................................................................

..................................................................................................................................................................

..................................................................................................................................................................

..................................................................................................................................................................

..................................................................................................................................................................

..................................................................................................................................................................

..................................................................................................................................................................

..................................................................................................................................................................

# 1. Appendix B

## Task 1: The 'Wall' Task

You are to develop a spreadsheet model to create a bid price for the total cost of building a garden wall. There will be two options for materials, lava rock, or brick.

The wall will be 6 meters long, 2 meters high, and 0.60 meters thick.

Crews of two workers will build both walls. Crews will work three eight-hour days to build either type of wall. Wages will $10.00 per hour per worker, plus an on cost of 20% to cover worker benefits. Lava rock costs $105.00 per cubic meter, and brick costs $70.00 per cubic meter. Your bid must add a profit margin of 30%.

Adapted from:
Panko, R.R. and Sprauge, R.H.J. "Hitting the Wall: Errors in Developing and Code-Inspecting a 'Simple' Spreadsheet Model" *Decision Support Systems,* (22) 1999, pp.337-353.

# Appendix C

## Task 2: The 'Ball'Task

You are to develop a spreadsheet model to estimate the total cost of materials for building a passenger-carrying balloon. The balloon will be perfectly spherical in **shape, and will be inflated** to a pressure of 1.4 atmospheres. The required volume of gas at this pressure to provide sufficient lift is 5,500 cubic meters.

The materials required will be fabric for the balloon wall, and gas for filling.

You are to estimate the total prices for two alternatives: an hydrogen balloon, and an helium balloon. Both hydrogen and helium gas are supplied and costed at a pressure of 150 atmospheres. Hydrogen gas costs $4,250.00 per cubic meter, and helium gas costs $7,580.00 per cubic meter, at this pressure.

The fabric for the balloon wall costs $25.00 per square meter.

There will be waste and losses in manufacture and filling. These allowances are 12.00% for wall fabric, and 3.00% for gas, over and above the calculated amounts.

The formulae for spherical area and volume are: A = 4 * π * r^2; $V = \frac{4 * \pi * r^3}{3}$

Where pi, π, equals 3.1416.

The formula describing the relative volumes and pressures of gasses is Boyle's Law, which states that: $P_1V_1 = P_2V_2$. That is: Pressure multiplied by Volume at state 1, is equal to Pressure multiplied by Volume at state 2, provided temperature is constant.